\documentclass{llncs}

\usepackage{graphicx}
\usepackage{soul, color}

 \newcommand{\NHe}[1]{\noindent\hspace*{#1em}}

\newcommand{\MSU}[1]{\mbox{\raisebox{0ex}{$\stackrel{\scriptstyle
#1}{\,\longrightarrow\,}$} }}

\def\CA{{\mathcal{A}}}

\def\CC{{\mathcal{C}}}

\def\CF{{\mathcal{F}}}
\def\CH{{\mathcal{H}}}
\def\CI{{\mathcal{I}}}
\def\CK{{\mathcal{K}}}
\def\CM{{\mathcal{M}}}
\def\CN{{\mathcal{N}}}
\def\CP{{\mathcal{P}}}
\def\CR{{\mathcal{R}}}
\def\CS{{\mathcal{S}}}
\def\CT{{\mathcal{T}}}
\def\CX{{\mathcal{X}}}
\def\CY{{\mathcal{Y}}}

\begin{document}

\title{How to Guarantee  Secrecy for  Cryptographic Protocols }

\author{Dani\`ele Beauquier and Fr\'ed\'eric Gauche}

\institute{LACL CNRS FRE 2673, Universit\'e Paris 12 Val de Marne\\
61 Avenue du G\'en\'eral de Gaulle F-94010 Cr\'eteil Cedex, France\\
\{beauquier, gauche\}@univ-paris12.fr}

\maketitle

\begin{abstract}
In this paper we propose  a general definition of secrecy for cryptographic protocols in
the Dolev-Yao model. We give a sufficient condition ensuring  secrecy for protocols where
rules have  encryption depth at most two, that is satisfied by almost all practical
protocols. The only allowed primitives in the class of protocols we consider are pairing
and encryption with atomic keys. Moreover, we describe an algorithm of practical interest
which transforms a cryptographic protocol into a secure one from the point of view of
secrecy, without changing its original goal with respect to secrecy of nonces and keys,
provided the protocol satisfies some conditions. These conditions are not very
restrictive and are satisfied for most practical protocols.\end{abstract}

 \section{Introduction}
Cryptographic protocols are used to ensure secure communications between two or more
parties in a distributed system. Among the requirements that cryptographic protocols must
satisfy are the well-known  authentication and secrecy or confidentiality.

Security protocol design and verification is a very hard problem. Sources of difficulty
are numerous and of different types. The  seminal paper for developing a model was
proposed by D. Dolev and A.C. Yao. \cite{DY83}. In recent years a lot of methods have
been proposed for reasoning about cryptographic protocols.  Some of them are based on the
trace model \cite{Pau98,GT02} including models with an explicit state-transition system
\cite{CDL+99} or Horn clauses \cite{Bla01,CCM01}. Another type of model uses processes to
represent cryptographic protocols \cite{AG97,Sch97}.

Concerning secrecy there are basically two approaches, the first one reduces the secrecy
property to a reachability problem, the second one defines secrecy in terms of an
observability equivalence.

Most of the papers are devoted to decidability and undecidability results depending on
various hypothesis related to the boundedness of nonces and sessions,  the
 used cryptographic primitives and so on. See for example \cite{DLM+04} for a review of these
 results. Surprisingly, there are  very few results that give some rules to apply in
 order to guarantee the secrecy property.
  This question has
 already been answered in the case of cryptographic  protocols using symmetric keys in
 \cite{BEA04},  which gives a sufficient condition for {solving} this problem. Here we consider
  a more general class of cryptographic protocols using  both symmetric and asymmetric
 keys. We give a new sufficient
 condition adapted to this type of protocols and we describe an algorithm of practical
 interest.
It transforms a cryptographic protocol (satisfying a condition which is not very
restrictive)
 into a secure one from the point of view of secrecy, without
changing its original goal with respect to secrecy of nonces and keys.
\newline In Section 2 we describe the model. Section 3 gives the sufficient condition
for a secure protocol w.r.t secrecy,  Section 4 is devoted to  the algorithm which
transforms a cryptographic protocol into a secure one w.r.t secrecy. The last section
concludes.\newline {\bf Related work} As mentioned before, the security literature
concentrates more on the verification of cryptographic protocols comparing  to the
synthesis of correct protocols. In \cite{AN96},  some prudent principles for designing
protocols are given but do not guarantee  the success. Several of these principles are
present in our definition of well composed protocol. A sufficient condition based on
typing is presented in \cite{abadi99secrecy} but it concerns only symmetric keys and a
binary view of secrecy according to which the world is divided into system and attacker.
Our sufficient condition can be considered as a generalization of the sufficient
condition given in \cite{LOW95}. Indeed, the protocols which are considered in this paper
do not admit forwarding, which is an important restriction. At last, to our knowledge,
there is no paper which describes an algorithm which transforms a protocol into a secure
one w.r.t. secrecy and preserves its original goal.
\section{The model}
In this section we formalize the model we use and we specify the assumptions we make
about protocols commonly referred to as the ``Dolev-Yao model''. Our approach is largely
inspired by \cite{Lowe}. The only primitives used are pairing and encryption. We assume
that pairing is \emph{associative}, which corresponds to practical protocols, so the
algebra of terms is the quotient of a free algebra  with equations for associativity.

We are interested in the behavior of the protocol when the number of agents, nonces and
sessions is \emph{unbounded}. Moreover the hypothesis on the power of honest agents is as
weak as possible. The knowledge of an agent is \emph{local}, it does not have a global
memory of all its sessions. On the contrary the power of the intruder is maximal.
\subsection{Messages}
\subsubsection{Atomic values}

 The set \emph{Value} is a set of disjoint types \emph{Agent}, \emph{Nonce},  \emph{Key},
 \emph{Cypher}.
\emph{Agent}, \emph{Nonce}, \emph{Key} are sets of atomic values. \emph{Cypher} is the
set of values obtained by encryption.
 The set \emph{Agent} is the set of all agent identities. It is partitioned into two subsets,
  honest agents and  intruders: \emph{$Agent = Honest \cup Dishonest$}. W.l.o.g. one supposes that the set $Dishonest$
  contains a unique intruder $\CI$.
 Agents variables, named A, B, ..., belong to the set \emph{AgVar}.
 \emph{Nonce} is an infinite set of integers.
 Nonce variables named \emph{N}, \emph{Na}, \emph{Nb}, ... , belong to the set \emph{NVar}.\newline
 The set \emph{Key} is divided into two disjoint subsets \emph{ShKey} and \emph{LKey}.\newline
 - \emph{ShKey} is the set of \emph{short term keys}. Its elements correspond to symmetric
  keys used only for the current session.\newline
 - \emph{LKey} is the set of \emph{long term keys}. It is a disjoint union of
 \emph{SymKey} the set of symmetric keys and \emph{AsymKey} the set of asymmetric keys.\newline
The set \emph{AsymKey} is a disjoint union of two subsets \emph{PubKey} (public keys) and
\emph{PrivKey} (private keys). \newline
 Short term key variables named $\CK$, $\CK'$, ... belong to the set \emph{KVar}.\newline
 Let $A, B$ be two agent variables. We denote respectively
by $\CK_{priv}(A)$, $\CK_{pub}(A)$, $\CK(A,B)$, the long term private encryption key of
$A$, the long term public encryption key of $A$, the long term symmetric key shared by
agents $A$ and $B$. Notice that in this
 notation $\CK_{priv}(A)$, $\CK_{pub}(A)$ are both encryption keys, and $\CK_{pub}(A)$
  is NOT the inverse key of $\CK_{priv}(A)$ and vice-versa.

 \subsubsection{Symbolic terms.}
 \emph{Symbolic terms} are constructed using \emph{pairing} and \emph{encryption}.\newline
 The pairing of terms $\CX$ and $\CY$ is the term $<\CX,\CY>$, the encryption of term $\CX$
  using the key $\CK$  is $\{\CX\}_{\CK}$, $C$ is a set of constants.\newline
 The grammar used to generate symbolic terms is :\newline
 $key ::= C\mid KVar \mid \CK_{pub}(AgVar) \mid \CK_{priv}(AgVar)\mid \CK(AgVar,AgVar)$\newline
 $symb\_term ::= NVar \mid AgVar \mid key \mid <symb\_term,symb\_term>
\mid \{symb\_term\}_{key}$\newline The pairing is associative, or equivalently for each
$n$ we have a primitive for $n$-pairing. We will consider only terms which are in a
"canonical form". For example the canonical form of terms $<t_1,<t_2,t_3>>$ and
$<<t_1,t_2>,t_3>$ is $<t_1,t_2,t_3>$, it means that $<t_1,<t_2,t_3>>$ and $<t_1,t_2,t_3>$
must be considered as triples and not pairs.

   The
set of subterms of a term $\tau$ is denoted $Sub(\tau)$.

 \subsubsection{Concrete terms.}
 {\emph{Concrete terms}} are generated following the same grammar as for symbolic terms,
  except that variables in $NVar, KVar, AgVar$ are replaced by  the value{s} of the corresponding type.\newline
  $Key ::= ShKey \mid \CK_{pub}(Agent) \mid \CK_{priv}(Agent)\mid \CK(Agent,Agent)$\newline
 $conc\_term ::= Nonce \mid Agent \mid Key \mid <conc\_term,conc\_term>
\mid \{conc\_term\}_{Key}$

 \subsubsection{Synthesis and Analysis.}
 In this subsection, 'term' means 'symbolic term'.
 The synthesis procedure represents  terms  the agents can build.
 The analysis procedure represents  terms  the agents can learn.\newline
 \newline
 Let $\CT$ be a set of terms and $A$ be an agent variable.
The set $Synth_A(\CT)$ is the least set of terms containing $\CT$ and satisfying:
 \begin{itemize}
 \item $\tau_1, ..., \tau_p \in Synth_A(\CT) \Rightarrow <\tau_1, ..., \tau_p> \in Synth_A(\CT)$ (An agent can
  compose the  terms he knows).
 \item $\forall \tau \in Synth_A(\CT), \forall B \in AgVar$
 \begin{itemize}
    \item  $\{\tau\}_{\CK(A,B)} \in Synth_A(\CT)$ (An agent can encrypt with  a  symmetric key he shares with another agent)
    \item $\{\tau\}_{\CK_{priv}(A)} \in Synth_A(\CT)$ (An agent can encrypt with his own private key).
    \item  $\{\tau\}_{\CK_{pub}(B)} \in Synth_A(\CT)$ (An agent can encrypt with the public key of any agent).
 \end{itemize}
 \item $\forall \tau \in Synth_A(\CT)$ and for all short term key variable  $\CK \in \CT$, $\{${$\tau$}$\}_{\CK} \in Synth_A(\CT)$ (An agent  can encrypt with all encrypting short term key he knows).
 \end{itemize}
 Let $A$ be an agent variable. Let $\CT$ and $\CT'$ be two sets of terms.\newline
 We have $\CT Anal_A \CT'$ if one of the following properties holds:
 \begin{itemize}
 \item $<\tau_1, ..., \tau_p> \in \CT, p>1$ and $\CT' = (\CT \setminus \{<\tau_1, ...,
 \tau_p>\}) \cup \{\tau_1\} \cup ... \cup  \{\tau_p\}$ (An agent can decompose terms).
 \item $\{\tau\}_{\CK} \in \CT$, $\CK \in \CT$ is a symmetric session key variable,
 and $\CT' = (\CT \setminus \{\tau\}_{\CK}) \cup \{\tau\}$  (An agent can decrypt terms
  encrypted with a short term session key he knows).
 \item $\{\tau\}_{\CK(A,B)} \in \CT$,
  $B \in \CT$ and $\CT' = (\CT  \setminus \{\tau\}_{\CK(A,B)})
  \cup \{\tau\}$ (An agent can decrypt terms encrypted with a  key shared with an agent  he knows.)
 \item $\{\tau\}_{\CK_{pub}(A)} \in \CT$ and $\CT' =
  (\CT \setminus \{\tau\}_{\CK_{pub}(A)}) \cup  \{\tau\}$ (An agent can decrypt terms
   encrypted with his own  public key).
  \item $\{\tau\}_{\CK_{priv}(B)} \in \CT$, $B\in \CT$ and $\CT' =
  (\CT \setminus \{\tau\}_{\CK_{priv}(B)}) \cup  \{\tau\}$ (An agent can decrypt terms encrypted with the private key of an agent he knows).
 \end{itemize}

 A set of terms $\CT$ is told \emph{undecomposable} if there does not exist any set of term $\CT'$ such that $\CT  Anal_A \CT'$ (An agent cannot decompose  any more term).\newline
 It is easy to prove that for any set of terms $\CT$, there exists a unique undecomposable set of terms $\CT'$ such that $\CT Anal_A ^*\CT'$. This  set is denoted $Anal_A^*(\CT)$.\newline
\newline
 For a term $\tau \in Anal_A^*(\CT)$ we define \emph{the number of steps necessary for $A$ to learn $\tau$ from $\CT$} as the number of decryption operations  that $A$ must use before obtaining $\tau$, more precisely :
 \begin{itemize}
 \item $A$ learns $\tau$ from $\CT$ in 0 step iff  some term $<...,\tau,...>$ is in $\CT$
  ( we admit
 here a composition of a single element $\tau$).
 (No  decryption necessary).
 \item if $A$ learns $\{<...,\tau,...>\}_{\CK}$ from $\CT$ in at most $p$ steps and
 $\CK$
 is a short session symmetric key learnt by $A$ from $\CT$ in at most $q$ steps, then
 $\tau$ is learnt by $A$ from $\CT$ in at most $p+q+1$ steps.
\item if $A$ learns $\{<...,\tau,...>\}_{\CK(A,B)}$ from $\CT$ in at most $p$ steps and
 $B$
 is  learnt by $A$ from $\CT$ in at most $q$ steps, then
 $\tau$ is learnt by $A$ from $\CT$ in at most $p+q+1$ steps.
\item if $A$ learns $\{<...,\tau,...>\}_{\CK_{pub}(A)}$ from $\CT$ in at most $p$ steps
 then
 $\tau$ is learnt by $A$ from $\CT$ in at most $p+1$ steps.
\item if $A$ learns $\{<...,\tau,...>\}_{\CK_{priv}(B)}$ from $\CT$ in at most $p$ steps
and
 $B$
 is  learnt by $A$ from $\CT$ in at most $q$ steps, then
 $\tau$ is learnt by $A$ from $\CT$ in at most $p+q+1$ steps.

 \end{itemize}
 For $\tau, \tau' \in Anal_A^*(\CT)$ we define that $A$ learns $\tau$ from $\CT$
 before $\tau'$ if $A$ learns $\tau$ in $p$ steps,  $\tau'$ in $p'$ steps and $p<p'$.\newline
Now, given a concrete agent $a$, we can define in the same way a relation $Anal_a$ on
finite sets of concrete terms  as well as  the other notions defined above, replacing
agents, and keys variables
 by values of the corresponding type.

 \subsubsection{Message ( Component, Protocol) -Template}
 A \emph{component template} is either a variable or an encrypted term. A \emph{message template} or \emph{t-message} is a tuple of the form $(A,B,\tau)$
 where $A$ an{d} $B$ are distinct variables of agents representing
 respectively the sender and the receiver and $\tau$ is a term representing the \emph{content} of the message.\newline
 A \emph{concrete message} is a tuple $(a,b,m)$ where  $a$ and $b$ are agent
  values and $m$ is a  concrete term. It corresponds to the informal usual notation
 $A\rightarrow B : m$.
\newline
 A \emph{protocol template} or simply \emph{protocol} is a sequence of message templates.
 \newline
 A \emph{role} in a protocol template is an agent variable appearing in this protocol.\newline
 Given a protocol $P$ with a set of roles $\CR$, a \emph{session template}
 $Ses_A$ for role $A \in \CR$ is the  subsequence of message templates
  of $P$ in which role $A$ is sender or receiver.
\\
Our running example will be the protocol TMN \cite{TMN90} using asymmetric keys. Brackets
for pairing
 are omitted as usual.

 \begin{example}.

\noindent
 $01 - A \rightarrow S : B,\{K_a\}_{\CK_{pub}(S)}$
\newline  $02 -
S \rightarrow B : B,A$\newline
 $03 - B \rightarrow S : A,\{K_b\}_{\CK_{pub}(S)}$\newline
 $04 - S \rightarrow A : B,\{K_b\}_{K_a}$
 \end{example}
 $Ses_S$ is the entire protocol.
 $Ses_A$ is the sequence:
\newline
$ A \rightarrow S : B,\{K_a\}_{\CK_{pub}(S)}$\newline
 $ S \rightarrow A : B,\{K_b\}_{K_a}$
 \subsection{Realizable protocol template}
 An elementary question is whether a protocol is ``realizable'', i.e.
 whether the honest agents can execute it. This notion appears in   \cite{RS03c} as "well-formed" protocol.
  We formalize this notion in our framework and give an algorithm which checks
 whether a protocol is realizable or not. One can observe that as far as we are
  aware of, most of the undecidability proofs
 \cite{DLMS:1999:UBSP,AC02,AmadioC02} are based on protocols which are not realizable, which
 is a weakness of these proofs. Only in \cite{CCM01} the undecidability proof relies on
 realizable protocols.\newline
 Let $P$ be a protocol, and $A$ be a role of this protocol. Consider the sequence of
 t-messages
 of the
 session template $Ses_A$. The $j^{th}$ t-message of $Ses_A$ is  of form $( A, B_j,
 \tau_j)$ or $( B_j,
 A,\tau_j)$ depending on  $A$ is sender or receiver of the message.\newline
 We define $Kn_{A,j}$ as the \emph{knowledge} of role $A$ after execution
 of message  number $j$. That is to say as the set  of terms known by $A$ after the execution of the first $j$ t-messages of his session and that $A$ can no more decompose.\newline
 This knowledge can be decomposed into two subsets:
 \begin{itemize}
 \item The \emph{basic knowledge} of $A$ at step $j$, $BasKn_{A,j}$, which contains agent, nonce and key variables.
 \item The \emph{cryptographic knowledge} of $A$ at step $j$, $CrKn_{A,j}$, which
 contains the encrypted terms known  by $A$ at step $j$ and he cannot decrypt.
 \end{itemize}
 Notice that $Kn_{A,j}$ contains only terms which are component templates.\newline
 From the definition of synthesis, we can define $Synth_A(Kn_{A,j})$ as the set of terms that $A$ can build from his  knowledge at step $j$.\newline
 Let us define by induction on $j$ the set $Kn_{A,j}$ and the fact that the $j$ first messages
 of $Ses_A$ are realizable.\newline
 The \emph{initial knowledge} of $A$, $Kn_{A,0}$ is fixed by the protocol.\newline
 We need to introduce the notion of new variables appearing in a t-message of protocol
 $P$. Let $(A_p,B_p,\tau_p)$ be the $p^{th}$ t-message of $P$. The set of new variables of this t-message denoted $NewVar_p$ is defined recursively:\newline
 $NewVar_1=Sub(\tau_1)\cap(AgVar\cup NVar\cup KVar)$.\newline
$NewVar_p=Sub(\tau_p)\cap(AgVar\cup NVar\cup KVar)\setminus(NewVar_1\cup...\cup
NewVar_{p-1})$ for  {$p$}$>1$.\newline
 Let $j>0$ and  suppose that the first $(j-1)$ messages
  are realizable by $A$ and  $Kn_{A,j-1}$ is defined, then:
 \begin{itemize}
 \item If in message number $j$, $A$ is receiver, this message
  can be realized by $A$ since $A$ is passive in this  action.
 \item If  message number $j$ is of the form $(A,B_j,\tau_j)$,
 this message can be realized by $A$
 if and only if: $\tau_j \in  Synth_A(Kn_{A,j-1}\cup NewVar_{p_j})$ where $p_j$
  is the index of the message $(A,B_j,\tau_j)$ in $P$.
 \end{itemize}
 In both cases, we have : $Kn_{A,j} = \{B_j\} \cup Anal_A^*(\{\tau_j\} \cup
  Kn_{A,j-1})$.
 \newline
 A session template $Ses_A$ is \emph{realizable} if all its t-messages in this session are realizable by role  $A$.\newline
 A protocol is \emph{realizable} if all the session templates of all roles of the protocol
 are realizable. Clearly, the above procedure is effective so one can decide whether
  a protocol is realizable.\newline
   For example on the TMN protocol with public
   key of the server, the evolution of the knowledge for each role is :\\[1ex]

 \begin{table}
\begin{tabular}{|c|c|c|c|}
\hline $ $     & A                & B                & S       \\ \hline
Initial &
$S,\CK_{pub}(S)$ & $S,\CK_{pub}(S)$ &    \\ \hline
 Step 1  & $B,\CK_a$        & $ $ &
$A,B,\CK_a$ \\ \hline
 Step 2  & $ $              & $A$              & $ $     \\ \hline Step
3  & $ $              & $\CK_b$          & $\CK_b$ \\ \hline
Step 4  & $\CK_b$          & $ $              & $     $ \\
\hline
\end{tabular}
\end{table}

 \textbf{By now, we will consider only realizable protocols.}
 \subsection{ States. Transitions}
 We formulate now the semantics of a protocol as an infinite transition system where a
 state contains the set of current partial sessions of agents (it is actually a multiset
  because the same
  agent may have several ``identical'' partial sessions  at the same time)  and a transition
 corresponds to a send or a receive event. As in \cite{LOW95} we assume that every message
 is intercepted by the intruder, so w.l.o.g. one consider that every sent message is sent to
 the intruder, and every received message is received from the intruder, so we have two
 types of events the \emph{send} and \emph{receive} ones.
  \subsubsection{States
  }
   A \emph{valuation} $v$ of a set of component template $\CT$ is a function
   that associates to each term $\tau \in \CT$  a concrete term $\bar{\tau} = v(\tau)$,
    the value of which is in $Value$
 (i.e. to each  component template is  associated its value). We consider here constants
  as variables for which the valuation is fixed.\newline
 Let $(\tau_j)_{j=1,...,k}$ be the list of contents of the t-messages
  of the session of a role $A$ for a protocol $P$. Let $v_j$ be a valuation for
  $Kn_{A,j}$. We denote $\tau_j[v_j]$ the concrete term we obtain when
  substituting in  term $\tau_j$ to each maximal subterm $\tau'$ which is in $Kn_{A,j}$
     the value  $v_j(\tau')$. One can remark that $\tau_j$ is built in a unique way from its
     maximal subterms which are in $Kn_{A,j}$.\newline
 A \emph{partial session} (or simply session) $\sigma$  is determined by its length $l$,   a role
  $A$ and a valuation $v_l(A)$ for the knowledge $Kn_{A,l}$. The role of session
   $\sigma$ will be denoted $R_{\sigma}$. The role $A$, the length $l$ and the valuation $v_l$
    permit to define the list of the $l$ first messages received by the agent playing
    this role in this session. It is the list of concrete messages $(\tau_j[v])_{j=1,  ...,
     p \leq k}$, where $(\tau_j)_{j=1, ..., k}$ is the list of the t-messages
     of the session of  role $A$.
\newline
 A \emph{state} is a multiset of partial sessions like in \cite{CDL+99}.

 \subsubsection{Transitions}
 The formalization of the evolution of the state of the system via receive or send events
 is the most delicate part of the modeling.
 An \emph{admissible state} is a state reachable from the initial state
 using transitions labeled by the following events:
 \begin{itemize}
 \item \emph{send event} : tuple $(a,\longrightarrow,(a,b,m))$ where  $a$, $b$
 are agents and  $(a,b,m)$ is a concrete message. It corresponds to the event "agent $a$ sends
 (intentionally to agent $b$) the message $m\,$ and this message is received by the intruder".
 \item \emph{receive event} : tuple $(a,\longleftarrow,(a,b,m))$ where  $a$, $b$ are agents and  $(a,b,m)$ is a concrete message. It corresponds to the event "The intruder
 sends to agent $b$ a message $m$  and agent $b$ believes that this message has been sent by agent $a$".
 \end{itemize}
 The \emph{knowledge of the intruder} denoted  $IntrKn$, is the set
 of values known by the intruder and that he cannot  decompose more. It
 will be described more precisely below.\newline
 $\bullet$ Send transitions\newline
 We have a transition from state $S$ to state $S'$ labeled by the send event
\\
  $(a,\rightarrow,(a,b,m))$, denoted by $S$ {\mbox{\raisebox{0ex}{$\stackrel{(a,
  \rightarrow,(a,b,m))}{\,\longrightarrow\,}$}}} $S'$ if  the following conditions are satisfied:
 \begin{enumerate}
 \item $a$, $b \in Agent$.
 \item There exists in $S$ a partial session $\sigma=(A,v_l)$ of length $l$
 for which the next message  is a  send event or agent $a$ starts a partial session for a role
 $A$ in which the first message of $Ses_A$ is a message sent by $A$.
 \item $(a,b,m) = \tau_{l+1}[v_{l+1}]$ where $v_{l+1}$ is a valuation defined as follows:
 \begin{enumerate}\itemindent 20pt
 \item $(v_{l+1} \mid BasKn_{R_{\sigma},l}) = (v_l \mid BasKn_{R_{\sigma},l})$
 \item $v_{l+1} \mid (BasKn_{R_{\sigma},l+1} \setminus BasKn_{R_{\sigma},l})$ must satisfy the rules
 \begin{itemize}\itemindent 10pt
 \item The values are of the correct type, i.e. values for nonces,
  agents and short term keys belong to the respective
 sets respectively $Nonce$, $Agent$, $Key$.
 \item The valuation is injective on the set of nonces and the set of keys, and
  values are ''fresh'', i.e., if $X$ is a variable for a nonce (resp. a key)
  belonging to $BasKn_{R_{\sigma},l+1} \setminus
  BasKn_{R_{\sigma},l}$, then $v_{l+1}(X)$ is not in the set of valuations of nonce
  variables (resp. key variables)
  for all the partial sessions of state $S$.
 \item $CrKn_{R_{\sigma},l+1} = CrKn_{R_{\sigma},l}$ and for coherence $(v_{l+1} \mid CrKn_{R_{\sigma},l+1}) = (v_l  \mid CrKn_{R_{\sigma},l+1})$.
 \end{itemize}
 \end{enumerate}
 \item $S'$ is the state we obtain when replacing one exemplary of session $\sigma=(l,A,v_l)$
 by $\sigma'=(l+1,A,v_{l+1})$.
  (It corresponds to increasing the list of concrete messages of the partial session $\sigma$ with the  concrete message $(a,b,m)$).
 \end{enumerate}
 The knowledge of the intruder $\CI$ at state  $S'$ is :

 $IntrKn_{S'} = Anal_{\CI}^*(IntrKn_S
 \cup \{(a,b,m)\})$\newline
 $\bullet$ Receive transitions\newline
 We consider here only the receive events where  the message is accepted by the receiver.\newline
 We have a transition {from state $S$ to state $S'$ labeled by the receive event
 $(a,\leftarrow~,(a,b,m)),$ denoted by} $S$
 {\mbox{\raisebox{0ex}{$\stackrel{(a,\leftarrow,(b,a,m))}{\,\longrightarrow\,}$}}} $S'$
 if the following conditions are satisfied: \begin{enumerate}
 \item $a$, $b \in Agent$.
 \item There exists in $S$ a partial session $\sigma=(l,A,v_l)$  for
  which the next message is a  receive event, or (case $l=0$) agent $a$ starts a partial session for a role
 $A$ in which the first message of $Ses_A$ is a message received by $A$.
 \item $(b,a,m) = \tau_{l+1}[v_{l+1}]$ where $v_{l+1}$ is a valuation defined as follows:
 \begin{enumerate}\itemindent 20pt
 \item $v_{l+1} \mid BasKn_{R_{\sigma},l} = v_l \mid BasKn_{R_{\sigma},l}$
 \item $v_{l+1} \mid (BasKn_{R_{\sigma},l+1} \setminus BasKn_{R_{\sigma},l})$ must satisfy the rules
 \begin{itemize}\itemindent 10pt
 \item  values belong to the set $Synth_{\CI}(S)$ defined above.
 \item  values of agent variables belong to $Agent$.
 \end{itemize}
 \item $v_{l+1} \mid (CrKn_{R_{\sigma},l+1} \cap CrKn_{R_{\sigma},l}) = v_l \mid (
 CrKn_{R_{\sigma},l+1} \cap
  CrKn_{R_{\sigma},l})$.
 \item $v_{l+1} \mid (CrKn_{R_{\sigma},l+1} \setminus CrKn_{R_{\sigma},l})$ has values in $Synth_{\CI}(S)$.
 \end{enumerate}
 \item $S'$ is the state we obtain when replacing
 an exemplary of partial session $\sigma$ with $\sigma'=(l+1,A,v_{l+1})$. (It corresponds
 to increasing the list of concrete messages of  the
 partial session $\sigma$ with the  concrete message $(b,a,m)$.
 \end{enumerate}
 The knowledge of the intruder $\CI$ at state  $S'$ is :

 {$IntrKn_{S'} = Anal_{\CI}^*(IntrKn_S \cup  \{m\})$}
 \newline
 The set $Synth_{\CI}(S)$ is the set of concrete terms that the intruder can build at state $S$.
 It is the least set  containing $IntrKn_S$ and satisfying:
 \begin{itemize}
 \item $\tau_1, ..., \tau_p \in Synth_{\CI}(S) \Rightarrow <\tau_1, ..., \tau_p> \in
 Synth_{\CI}(S)$.
 \item {$Agent\subset Synth_{\CI}(S)$.}
 \item {For every agent $a$, the long term key $\CK(a,\CI)$ is in $Synth_{\CI}(S)$.}
 \item {For every agent $a$, the long term key $\CK_{pub}(a)$ is in $Synth_{\CI}(S)$.}
\item For every term $\tau \in Synth_{\CI}(S)$ and for every key $\CK \in (IntrKn_S \cap
Key)$, $\{\tau\}_{\CK} \in Synth_{\CI}(S)$. \end{itemize}
 A \emph{trace} of a protocol is a sequence $S_0$
 {\mbox{\raisebox{0ex}{$\stackrel{e_1}{\,\longrightarrow\,}$}}} $S_1$
   {\mbox{\raisebox{0ex}{$\stackrel{e_2}{\,\longrightarrow\,}$}}} ...$S_{n-1}$
    {\mbox{\raisebox{0ex}{$\stackrel{e_n}{\,\longrightarrow\,}$}}} $S_n$ where
    $S_0$ is the initial state and each $S_{i-1}\MSU{e_i} S_i$ is a transition.
\\
The  initial knowledge of the intruder $IntrKn_{S_0}$ is given by the protocol.
\\[1ex]
{\bf Remark.} One can notice that the rules applied by an honest agent in order to accept
a message correspond to a very weak control of the message. The agent makes only
\emph{equality} tests, it  has no possibility to control for example the depth of
encryption, the correct type  of values and so on.
 \subsection{Secrecy}
 In the literature, generally the definitions of secrecy   are
 very dependent  on the chosen model and
 restrictive, i.e. sufficient for the hypothesis made by the authors but not applicable in
 a more general context. The definition we give here seems very general,
 at least as far as the concern is the secrecy of values and not of properties.

 \begin{definition} The \emph{secret} of a variable $X$ for a nonce or a short term key can be broken from the
 point of view of A if there exists  a reachable state $S$ containing a partial session
 $\sigma$ of length $l$ for role $A$ with valuation $v_l$ for $Kn_{A,l}$ such that
 \begin{enumerate}
    \item $BasKn_{A,l}$ contains $X$ and the set $\CR$ of  roles of the protocol
    \item $\CI$ does
 not belong to the valuation $v_l(\CR)$  ($\CI$ does not participate to the partial
 session $\sigma$
 from the point of view of $A$)
    \item $v_l(X) \in  IntrKn_S$.
 \end{enumerate}
 \end{definition}
 As one can observe, the notion of secrecy implies two parameters: a \emph{variable} for
 which the secret is broken and a \emph{role} which can claim the fact. We have to justify
 points 1 and 2. Why should the set $\CR$  be in $BasKn_{A,l}$? Because as far as the agent
 involved in
   the partial session $\sigma$ does not know all its partners in this session, it cannot
   claim whether it is correct that the agent $\CI$ knows the value $v_l(X)$. Indeed, if
   $\CI$ participates in an honest way to the session it is normal that $v_l(X) \in
   IntrKn_S$. For the same reason the condition that $\CI$ does
 not belong to the valuation $v_l(\CR)$ is required. An unsolved question is how to
 define secrecy in the case when the set of roles does not belong to the knowledge of
 each role at the end of its partial session.
\\
 There is a well-known attack \cite{LR97} on the  protocol TMN of Example 1. An intruder $\CI_A$ acts as
 if it was $A$:
\newline
$01 - \CI_a \rightarrow S : <b,\{K_i\}_{\CK_{pub}(S)}>$
\newline  $02 -
S \rightarrow b : <a,b>$\newline
 $03 - b \rightarrow S : <a,\{K_b\}_{\CK_{pub}(S)}>$\newline
 $04 - S \rightarrow \CI_a : <B,\{K_b\}_{K_i}>$\newline
 In this attack, the secret is broken for the variable $K_b$ from the point of view
 of $B$ because the trace given here reaches a state containing a partial session for
 role $B$ satisfying the above  three conditions.
\\[2ex]
 Given a protocol, the variables which can be learnt by an external observer
  of the protocol are called \emph{revealed  variables}. The others
  (those which remain unaccessible to this observer) are called \emph{unrevealed  variables}.\newline
 More precisely, given a protocol  $P = (A_i,B_i,M_i)_{i=1, ..., k}$,  a variable $X$
  for a nonce or a key is revealed in $P$ if $X \in Anal^*_C(\{M_1,  ..., M_k\})$ for some  $C$
   not being a role of $P$. The set of revealed variables of a protocol is clearly computable.
   In an obvious way, the secret can be broken for every
   revealed variable from the point of view of every role. Thus, the interesting
    question is ``can the secret be broken for an unrevealed variable''. The next section
    answers to this question by giving a sufficient condition which guarantees that
    the protocol preserves the secrecy of unrevealed variables for nonces and
 short {term} key variables.

\section{A sufficient condition for secrecy}
\subsection{Well-Composed Protocol}
 A \emph{signature} of a proto{c}ol is constituted by a
 nonce variable which is called the session nonce   and a fixed  list of the agent
 roles $<n,A_1,...,A_p>$.
 \begin{definition}\label{DefinitionWellComposed}
 A protocol  is \emph{well composed} if :
 \begin{enumerate}
 \item Encryption is of depth at most two.
 \item Private long term {asymmetric} {and long
 term symmetric keys} are never transmitted.
 \item There exists a signature $\CS$ such that \begin{description}
    \item[-]  the content of every t-message is a
 term of the form : $<{\cal S},\{{\cal S},m\}_{\CK_{priv}(A)}>$ where $A$ is the sender
 of the message,
    \item[-]  every subterm of the protocol which is an  encrypted term has the form $\{{<\cal
 S},...>\}_{\CK}$ (it contains the signature on the left inside the encryption).
 \end{description}
  \item Two different encrypted terms which are encrypted by the same type
 of keys (public, private, ...) must have a different number of elements. More precisely,
 if $\{<\tau_1,...\tau_k>\}_\CK$ and $\{<\tau'_1,...,\tau'_{k'}>\}_{\CK'}$ are two
 different subterms of a protocol $P$ and $\CK,\CK'$ are of the same type, then $\CK \not= \CK'$.
 \end{enumerate}
 \end{definition}
\noindent Let us comment the four given conditions. Condition (4) helps to prevent the
intruder from passing off  a term $\{\tau\}_\CK$ as a term $\{\tau'\}_{\CK'}$ while these
terms are intended  to be distinct terms in the specification. Another way to obtain the
same effect would be to use tagging as it is done in several papers
\cite{BP03,HLS00,RS03c}. In these papers, tagging is used to prove decidability of
secrecy for tagged protocols, but it is not a sufficient condition for secrecy. Condition
(3) is reasonable and permits to know at each moment who is supposed to be implied in the
session. An attack on TMN protocol is due to the fact that this condition is not
satisfied. Condition (2) is always recommended \cite{AN96}. At last, condition (1) is not
essential here. We are convinced
that this hypothesis could be relaxed, but it would make the proof more complicated.\\
 The TMN protocol  is not well composed. Here is a modified version which is well composed:\newline $01 - A \rightarrow
S : \CS,\{\CS,B,\{\CS,K_a\}_{\CK_{pub}(S)}\}_{K_{priv}(A)}$
\newline  $02 -
S \rightarrow B :\CS,\{\CS, B,A\}_{K_{priv}(S)}$\newline
 $03 - B \rightarrow S :\CS,\{\CS, A,\{\CS,K_b\}_{\CK_{pub}(S)}\}_{K_{priv}(B)}$\newline
 $04 - S \rightarrow A :\CS,\{\CS, B,\{\CS,K_b\}_{K_a}\}_{K_{priv}(S)}$\\[1ex]
 The attack presented in the previous section fails in this new
 version because the intruder cannot impersonate  $A$ at the first step of
 the attack.
\begin{theorem}\label{TheoremPreservationSecrecy}
 A well composed protocol preserves the secrecy of unrevealed variables for nonces and
 short {term} key variables.
\end{theorem}
\noindent Before giving the proof of this theorem let us recall the sufficient condition
given in \cite{BEA04}  to preserve secrecy in case of symmetric encryption, and show with
a counter example that this condition is not enough for protocols involving asymmetric
encryption. This sufficient condition was:
\begin{enumerate}
 \item Encryption is of depth  one.
 \item  Long
 term  keys are never transmitted.
 \item There exists a signature $\CS$ such that
     every subterm of the protocol which is an  encrypted term has the form $\{{<\cal
 S},...>\}_{\CK}$ (it contains the signature on the left inside the encryption).
\end{enumerate}
Here is a  variant of TMN protocol which satisfies this condition.
\newline
Let $\CS=<N,A,B>$ where $N$ is a nonce.\\[1ex]
\newline
$01 - A \rightarrow S : \CS,B,\{\CS,K_a\}_{\CK_{pub}(S)}$
\newline  $02 -
S \rightarrow B : \CS,B,A$\newline
 $03 - B \rightarrow S : \CS,A,\{\CS,K_b\}_{\CK_{pub}(S)}$\newline
 $04 - S \rightarrow A : \CS,B,\{\CS,K_b\}_{K_a}$.\\[1ex]
 Clearly, an attack similar to the one given before can be repeated.
\\
The next proposition expresses the fact that a well composed protocol guarantees some
authenticity: if an agent $a$ receives
 in a partial session where it plays role $A$ a message $m$  from another agent $b$ and $a$
 thinks that $b$ plays role $B$ and that $m$ corresponds to the message number $i$ of
 the protocol, indeed  $b$ has sent this message for this purpose.
\begin{proposition}\label{PropositionAssurance}
Let $r$ be a trace of  a well composed protocol. If $r$ contains a transition
$S\MSU{(a,\leftarrow,(b,a,\tau))}S'$ where $S$ contains a partial session $\sigma$ of
length $l$ belonging to an agent $a$ for the role $A$, and $\sigma$ is replaced in $S'$
by a partial session $\sigma'$ of length $l+1$ where $b$ has role $B$, then there is a
previous transition in $r$ of the form $S_1\MSU{(b,\rightarrow,(b,a,\tau))}S_1'$ where
$S$ contains a partial session $\sigma_1$ of length $l_1$ belonging to  agent $b$ for the
role $B$  and $\sigma_1$ is replaced in $S_1'$ by a partial session $\sigma_1'$ of length
$l_1+1$ where the message number $l+1$ of role $A$ is exactly the message number $l_1+1$
of role $B$.
\end{proposition}

 \begin{proof}
 If $a$ accepts the message, it means that the message is of the right form, namely :
 $(b,a,\tau)$ with $\tau=<s_1,\{s_1,\tau'\}_{\CK_{priv}(b)}>$.\newline
 Actually $\tau$ must be encrypted by $\CK_{priv}(b)$ since it is supposed to {have}
 be{en} sent by $b$. Moreover, $a$ {controls}  that
 the signature
 located in the first elements of $\tau$ is the same as the signature  contained
 at the beginning of the encrypted element. As a consequence,
 $b$ is the agent who encrypted $\tau$. {D}ue to the
 last condition of the definition of a well composed protocol, {$a$  also controls
 that} the number of elements in $\tau$ corresponds to the number of elements
 {a}waited {{by $a$}} in this session, so necessarily, $b$ built $\tau$
 to send a message number $l+1$ for the role $A$, and this role is played by $a$ because
 $a$ has in the signature the place corresponding to role $A$.
 \end{proof}
We now translate in an equivalent form the property of secrecy for a well composed
protocol.
 \noindent Let $r$ be a run with a length $l$, $X$ be an unrevealed variable for
a nonce or a short key, $x$ be a value, $T$ be a time less than or equal to $l$, and $t$
be a positive integer. The tuple $(r,X,x,T,t)$ satisfies $\CP_1$ (resp. $\CP_2$) iff:
\begin{itemize}
\item $\CP_1$: in $r$ at some time $T'<T$, in one of its  partial sessions whose
signature does not contain $\CI$, an honest agent $a$ generates the value $x$ to assign
to the unrevealed variable $X$ and at time $T$, $\CI$ learns the value $x$ in $t$ steps.
\item $\CP_2$: in $r$,  in one of its  partial sessions whose signature does not contain
$\CI$,
 an honest agent $a$ learns the value $x$ of the unrevealed variable $X$ at time $T$ in $t$ steps
 and at the end of the run $r$ the value $x$ belongs to the
knowledge of $\CI$, i.e. $x\in IntrKn_{l}$. Moreover, there is no tuple of the form
$(r,X',x,T',t')$ satisfying  $\CP_1$, in other words $x$ is not a value generated by an
honest agent to assign to an unrevealed variable.
\end{itemize}

\begin{lemma}\label{elem}
A well composed protocol preserves the secrecy of unrevealed variables for nonces and
short {term} key variables from the point of view of every role iff there does not exist
an unrevealed variable $X$ for a nonce or a short {term} key, a value $x$, a run $r$ with
length $l$, a time $T \leq l$ and a positive integer $t$ such that the tuple
$(r,X,x,T,t)$ satisfies $\CP_1\vee \CP_2$.
\end{lemma}
\begin{proof}
\noindent Firstly assume that there  exists an unrevealed variable $X$ for a nonce or a
short {term} key, a value $x$, a run $r$ with length $l$, a time $T \leq l$ and a
positive integer $t$ such that the tuple $(r,X,x,T,t)$ satisfies $\CP_1\vee \CP_2$.

If $(r,X,x,T,t)$ satisfies $\CP_1$ then in $r$ at some time $T'<T$, in some partial
session $\sigma$ for a role $A$, with a signature that does not contain $\CI$, an honest
agent $a$ generates the value $x$ to assign to the unrevealed variable $X$ and at time
$T$,in some state $S$, $\CI$ learns the value $x$ in $t$ steps. Clearly  the {secret} of
variable $X$ can be broken from the point of view of  role A. Actually in state $S$, the
extension of partial session $\sigma$ has a length $l$ and  a valuation $v_l$ for
$Kn_{A,l}$
 such that $BasKn_{A,l}$ contains $X$ and the set $\CR$ of roles of the protocol,
   $\CI$ does not belong to the valuation $v_{l}(\CR)$
 and $v_{l}(X) \in  IntrKn_S$.

 If $(r,X,x,T,t)$ satisfies $\CP_2$, in the same way let $A$ be the role played by $a$ in
 its partial session.  The {secret} of
variable $X$ is broken from the point of view of  role A. The ``if'' part of the Lemma is
proved.

  Secondly assume that in a well composed protocol, the {secret} of a
variable $X$ can be broken from the point of view of a role A. It means  there exists  a
reachable state $S$ containing a partial session $\sigma$ of length $l'$ for role $A$
 with valuation $v_{l'}$ for $Kn_{A,l'}$
 such that $BasKn_{A,l'}$ contains $X$ and the set $\CR$ of roles of the protocol,
   $\CI$ does not belong to the valuation $v_{l'}(\CR)$
 and $v_{l'}(X) \in  IntrKn_S$.
 Let $r$ be a run from the initial state of the protocol to state $S$,  let $l$ be its
 length and let $x=v_{l'}(X)$. Since $BasKn_{A,l'}$ contains $X$ it means that at some
 moment in the partial session $\sigma$ the agent $a=v_{l'}(A)$ either generates the
 value $x$ to assign to the variable $X$ (first case) or $a$ learns it (second case).

 In the first case, let $T'$ be the moment when $a$ generates the
 value $x$. Since $v_{l'}(X) \in  IntrKn_S$, there is a time $T>T'$ when the intruder
 learns $x$ in $t$ steps, more precisely, if $S_i$ denotes the $i$-th state of run $r$, there is a
 state $S_T$ such that $x\in IntrKn_{S_T}$ and $x\not\in IntrKn_{S_{T-1}}$. In this first
 case the tuple $(r,X,x,T,t)$ satisfies $\CP_1$.

 In the second case, in the partial session $\sigma$, $a$ has not generated $x$ (may be
 $a$ has generated $x$ in another session)  and $a$ has learnt the value $x$ of $X$ at time
 $T\leq l$ in $t$ steps. If there
 exists a tuple $(r,X',x,T'',t')$ satisfying $\CP_1$, we are done. If not, then the tuple
$(r,X,x,T,t)$ satisfies $\CP_2$. The ``only if'' part of the Lemma is proved.
\end{proof}

Well composed protocols have an invariant property which is stated below not very
formally:
\begin{lemma}\label{LemmaOnlyOccurencesEncryptedComponents}
If in a trace $r$, at time $T_1$, an honest agent $a$ generates a value $x$ to substitute
to an unrevealed variable $X$ in a message $m$ that he sends with a signature $\CS$ not
including $\CI$, then, as long as $\CI$ does not learn $x$, $x$ has only occurrences in
encrypted components $\tau = \{\CS,\dots,x,\dots\}_K$ where the term $\tau$ has been
encrypted by an honest agent belonging to $\CS$ and put by this same agent in a message
$m'$ in which the place where is $x$ inside $\tau$ is the place of $X$.
\end{lemma}

 \begin{proof}
The property is true at $t_1$. {L}et $t> t_1$ and assume $\CI$ does not know
 $x$ at $t$. If $x$ is in an encrypted component, this one has been encrypted  by an
 honest agent $b$, in some session  otherwise, $\CI$ knows $x$. The value $x$ is by
 recurrence hypothesis for $b$ the value of an unrevealed variable $X$, and then in the
 component encrypted by $b$ to send in a message $m'$, $x$ is in place of $X$.

 \end{proof}

\begin{proposition}\label{PropositionNoP1P2}
 In a well composed protocol, there does not exist any tuple satisfying $\CP_1 \vee \CP_2$.
\end{proposition}
\begin{proof}\noindent
 Suppose there exist tuples $(r,X,x,T,t)$ for which $\CP_1\vee \CP_2$ holds. Consider the total strict order relation: $(r,X,x,T,t)<(r',X',x',T',t')$ iff $T<T'$ or ($T=T'$ and $t<t'$) and take a minimal tuple $(r,X,x,T,t)$ satisfying $\CP_1\vee \CP_2$. Let us examine the two cases :
 \begin{itemize}
 \item The tuple  $(r,X,x,T,t)$ satisfies $\CP_1$.

Let $\CS$ be the signature not including $\CI$ of the partial session in which at time
$T_1<T$, the honest agent $a$ generates the value $x$ to substitute to the
 unrevealed variable $X$  in a message $m$.
  Let $\tau$ be the concrete term from which $\CI$ learns $x$ at time $T$ in $t$ steps.
 We consider here the term of the very last operation of decryption made by $\CI$ to
 learn $x$. This term  $\tau$ has a value of type $Cypher$ and
  it is of the form
  $\{...,x,...\}_{\CK}$.
   This term has not been  built by $\CI$ in $r$ before, otherwise $\CI$ would have known $x$
  before, and $(r,X,x,T,t)$ would not be minimal. So it has been built
  by an honest agent $d$, and for this reason{,} due to
  Lemma \ref{LemmaOnlyOccurencesEncryptedComponents}, it has been built by an honest
  agent $d$ belonging to $\CS$ and put by this same agent in a message $m'$ in which the
  place where is $x$ inside $\tau$ is the place of $X$. Thus the term $\tau$ is of the
  form  $\{\CS,...,x,...\}_{\CK}$, because the places of the unrevealed variable $X$
  cannot be the places of the components of $\CS$.
 There are 3 cases for $\CK$ :
 \begin{enumerate}
 \item  $\CK$ is a long term symmetric key
 \item  $\CK$ is a public long term key
 \item  $\CK$ is a short term symmetric key.
 \end{enumerate}
 Actually, $\CK$ cannot be a private key, because this private key should
  be $\CK_{priv}(d)$ and
{ the component would not be in a unrevealed position}. Let's go through each of the
three cases :
 \begin{enumerate}
 \item Since $d$ has built the term $\tau$, and since the signature inside $\tau$ does not contain $\CI$, $\CK$ is equal to some $\CK(c,d)$ where $c$ is a honest agent. So $\CI$ cannot decrypt $\tau$.
 This first case is not possible.
 \item The key $\CK$ cannot be the public key of $\CI$, because $\CI$ is not in the signature $\CS$. So $\CI$ cannot decrypt $\tau$.
 This second case is also impossible.
 \item
In the partial session $s$ where $d$ builds the term $\tau$ at time $T'<T$, either $d$
knows $\CK$ or he generates it. In both cases in the message $m'$ sent by $d$ which
contains $\tau$, $\CK$ is in place of an unrevealed variable $Y$, otherwise, $x$ itself
would be in an unrevealed place. Thus the secret is broken from the point of view of  the
role played by $d$ in this partial  session $s$ and for the variable $Y$ in the place of
$\CK$ in the term $\tau$ inside the message $m'$. As for $\CK$ there are two
possibilities. Either $d$ has generated it or he {has} learnt it at time at most $T'$ in
in this session $s$. In the first case, there is a  tuple $(r,Y,\CK, T",t')$ which
satisfies $\CP_1$ with $T"<T$. In the second case there is a  tuple $(r,Y,\CK, T",t')$
which satisfies $\CP_2$ with $T"<T$.
 It contradicts the minimality of $(r, X,x,T,t)$.
 \end{enumerate}

\item The tuple  $(r,X,x,T,t)$ satisfies $\CP_2$.

 At time $T$, in $t$ steps an honest agent $a$ in a
partial session $s$ whose signature $\CS$ does not contain $\CI$ learns the value $x$.
Let $\tau$ be the last encrypted concrete term from which $a$ learns $x$. This term
$\tau$ was contained in a message $m_1$ received by $a$ at time $T$ or before and $x$ in
this message $m_1$ is in place of an unrevealed variable $X$. This message which has been
accepted by $a$ has the form $<\CS,\{\CS,...,\tau,...\}_{K_{priv}(a_1)}>$ where $a_1\in
\CS$ or $<\CS,\tau>$. Here we use the fact that the protocol has an encryption depth at
most two. If $m_1$ was equal to $<\CS,\tau>$, then we would have
$\tau=\{..,x,.\}_{K_{priv}(c)}$, and $x$ would be in place of a revealed variable. So the
message has the form $<\CS,\{\CS,...,\tau,...\}_{K_{priv(a_1)}}>$. Moreover the term
$\tau$ has the form $\{...,x,...\}_{\CK}$. Let us observe that, the number of components
of $<\CS,\{\CS,...,\tau,...\}_{K_{priv(a_1)}}>$ permits to the agent $a$ to identify the
index $i$ of the message. For the same reason since the term
$\{\CS,...,\tau,...\}_{K_{priv}(a_1)}$ has been built by $a_1$ in some partial session
$s_1$, $a_1$ has built this term in order to send the  message $m_1$, and the value of
$\tau$ in the partial session $s_1$ of $a_1$ and in the partial session $s$ of $a$ are
associated to the same symbolic term of the protocol. Let us come back to
$\tau=\{...,x,...\}_{\CK}$.

 If the agent $a_1$ in the partial session $s_1$ builds the term $\tau$ by encryption
 with $\CK$ before sending the message $m_1$, it means that in this session $s_1$ at this
 mome{n}t, $x$ is known by $a_1$. So in this session $s_1$, $x$ is learnt by $a_1$ at
 a time $T'<T$ in $t'$ steps. Indeed, $x$ is  not generated by $a_1$ at least in this
 session because for $a_1$, in this session, $x$ is the value of $X$ which is unrevealed,
 which would contradict the fact that  $(r,X,x,T,t)$ satisfies $\CP_2$. Thus replacing
 $a$ by $a_1$ we get a tuple  $(r,X,x,T',t')$ satisfying $\CP_2$ {which} contradicts
 the minimality of $(r,X,x,T,t)$. So the agent $a_1$ in the partial session $s_1$ does
 not build the term $\tau$ by encryption with $\CK$ before sending the message $m_1$. It
 means that this term $\tau$ has been obtained from a previous message $m_2$ that the
 agent $a_1$ received in its partial session $s_1$. Thus  we can iterate our reasoning
 for $a_1$ instead of $a$, but only a finite number of times because the run $r$ is
 finite. Thus we get in any case a contradiction.
 \end{itemize}\noindent
 So we've proved that there cannot exist any tuple satisfying $\CP_1 \vee \CP_2 $, which
 by induction, proves Proposition \ref{PropositionNoP1P2}.\newline

\end{proof}
 Theorem \ref{TheoremPreservationSecrecy} is a direct consequence of Lemma
 \ref{elem} and {P}roposition \ref{PropositionNoP1P2}.\\[1ex]
A well composed version of TMN protocol would be:
\begin{example}.

\noindent
 $01 - A \rightarrow
S : \CS,\{\CS,B,\{\CS,K_a\}_{\CK_{pub}(S)}\}_{K_{priv}(A)}$
\newline
$02 - S \rightarrow B :\CS,\{\CS, B^2,A\}_{K_{priv}(S)}$\newline
 $03 - B \rightarrow S :\CS,\{\CS,B^2, A,\{\CS,B,K_b\}_{\CK_{pub}(S)}\}_{K_{priv}(B)}$\newline
 $04 - S \rightarrow A :\CS,\{\CS, B^5,\{\CS,K_b\}_{K_a}\}_{K_{priv}(S)}$

 ($B^n$ means a sequence of $n$ $B$)
 \end{example}
 The previous attack fails at the first step because the intruder cannot impersonate the
 agent $a$ for role $A$ in the first message of the protocol. There are other attacks
  on this protocol which use the algebraic properties of the XOR algorithm used for
  encryption, but it is out of the scope of our framework.
 \\[1ex]
 {\bf Remark.} As noticed by M. Abadi in \cite{abadi99secrecy}, authenticity is dual to
 secrecy in the sense that authenticity concerns the source of the messages while secrecy
 concerns their destination. Nevertheless, it seems that it is hard to ensure secrecy
 without some phase of authentication. If we look at  attacks that breach the secrecy
 without using
 specific algebraic properties of the  encryption algorithms, very often the intruder
 exploits
 some  weakness of the protocol with respect to authentication.
\section{An algorithm for securing protocols}
In this section we describe a very simple  algorithm $\CA$ which transforms a protocol
$P$ into a protocol  $P'=\CA(P)$ which is secure w.r.t{.} secrecy and such that $P'$
preserves the "intended goal" of $P$ for a large class $C$ of protocols. Surely, we have
to define what means "to preserve the intended goal". We first describe the class $C$,
secondly we give the algorithm $\CA$ which can be applied to every protocol in the class
$C$, then we define an equivalence relation over the set of protocols in $C$. Finally we
prove that for every protocol $P$  in $C$ the protocol $\CA(P)$ is equivalent to $P$ and
is well composed, so, $\CA(P)$ is secure w.r.t{.} secrecy.

\subsection{The class $C$ and the algorithm $\CA$}

\begin{definition}
A protocol is in $C$ if it satisfies the following conditions:
\begin{itemize}
\item Encryption is of depth at most two. \item If in a template message $(A,B,\tau)$
there is a subterm of $\tau$ with an encryption depth equal to two, then
$\tau'=\{\tau''\}_{K_{priv}(A)}$.
 \item Private long term {asymmetric}  {and long term
symmetric keys} are {never} transmitted.
\end{itemize}
\end{definition}
A lot of  protocols belong to the class $C$ :
 ISO/IEC 11770-3
Key Transport Mechanisms (1,2,3,4,5,6),  Helsinki Protocol, TMN with public key protocol,
Blake-Wilson-Menezes Secure Key Transport Protocol, Needham-Schroeder Public Key Protocol
X.509 one-pass, two-pass, three-pass authentication, \dots

\noindent The algorithm $\CA$ is the following :
\begin{enumerate}
\item  Introduce  a new variable $N$ for a  session nonce, and define a signature $\CS =
<N,R_1,R_2,...,R_k>$ where $R_1,R_2,...,R_k$ are the roles of the protocol $P$. \item
Transform the content $m$ of each template message $(A,B,m)$ according to the type of $m$
:
\begin{description}
\item[*] If $m$ is a tuple of $n$ elements ($n \geq 1$) and none of them is encrypted by
$K_{priv}(A)$, replace $m$ with $\{m\}_{K_{priv}(A)}$. \item[*] If
$m=<\tau_1,...,\tau_n>$  ($n\geq 1$) and at least one of the $\tau_i$ is encrypted by
$K_{priv}(A)$, replace $m$ with $\{m'\}_{K_{priv}(A)}$ where $m'$ is the term we get,
replacing each term $\tau_i=\{\tau'_i\}_{K_{priv}(A)}$ with $\tau'_i$. In other terms,
the encryption with $K_{priv}(A)$ is done over the tuple instead of some of its elements.
\end{description}
\item Replace in each template message, each subterm of the form $\{\tau\}_K$ by the
subterm $\{<\CS,\tau>\}_K$. Notice that, by associativity we have
$<\CS,\tau>=<N,R_1,R_2,...,R_k,\tau>$ \item Replace each content $m$ with $<\CS,m>$.
\item If several terms of the protocol encrypted by the same type of key namely long term
public type, long term private type, long term symmetric type or short term symmetric
type have the same number of elements, add inside the term, after the signature,
occurrences of the last role in order to get different numbers of elements for all the
encrypted terms of the same type.
\end{enumerate}
The well composed protocol of Example 2 is obtained  applying this algorithm to Example
1.

 \noindent We
now prove that the protocol $P'$ one obtains  applying the algorithm $\CA$ to a protocol
$P \in C$ is in some sense equivalent to $P$, i.e. the new knowledge of each role is
essentially the same as before, at least from the point of view of the nonces and the
session keys appearing in the protocol $P$.

\begin{definition}
Let $P$ be a protocol in the class $C$ and $P' = \CA(P)$. The protocol $P'$ is said
\emph{weakly equivalent} to $P$ if for each role $R_i$ for each step $j$,
$BasKn_{R_i,j}(P') = BasKn_{R_i,j}(P) \cup \{R_1,...,R_n\} \cup \{n\}$ {and}
$CrKn_{R_i,j}(P') = \sigma(CrKn_{R_i,j}(P))$ where the $\sigma(\{\tau\}_k) =
\{<\CS,\tau>\}_k$ for every term $\tau$. (Notice that terms of $CrKn_{R_i,j}$ have an
encryption depth equal to 1 for protocols in the class $C$).
\end{definition}
In other terms, at every step, the basic knowledge is only increased  by the set of roles
and the nonce which is added, and the encrypted knowledge is the same except that the
signature in inserted in the encrypted term.

\begin{theorem}
Let $P$ be a protocol in the class $C$ and $P' = \CA(P)$. The protocol $P'$ is
\emph{weakly equivalent} to $P$ and is well composed.
\end{theorem}

\begin{proof}
Recall that for every role $A$, $Kn_{A,j}(P) = \{B_j\} \cup Anal_A^*(\{\tau_j\} \cup
Kn_{A,j-1})$ for role $A$ if the $j$-th template message of his partial session is
$(A,B_j,\tau_j)$ or $(B_j,A,\tau_j)$.

 Let $(A,B_j,<\CS,\{\CS,\tau'_j\}_{K_{priv}(A)}>)$
resp. $(B_j,A, <\CS,\{\CS,\tau'_j\}_{K_{priv}(B_j)}>)$ {be} the corresponding message in
$P'$. We have

$Kn_{A,j}(P')=\{B_j\}\cup Anal_A^*(<\CS,\{\CS,\tau'_j\}_{K_{priv}(A)}>)\cup
Kn_{A,j-1}(P')$,

 where $\tau'$ is obtained from $\tau$ essentially by adding the
signature in every encrypted term. So by induction on $j$,

$Kn_{A,j}(P')=BasKn_{A,j}(P)\cup \{R_1,...,R_n\}\cup \{N\}\cup \sigma(CrKn_{A,j}(P))$

where $\CS=<N,R_1,...,R_n>$.
\end{proof}
{\bf Remark} The condition concerning the number of elements inside encrypted
 terms of the same type can be obtained more simply by adding different integers inside the
 encrypted terms which permit to identify them. Proceeding in this way,
 the messages will be shorter.
\section{Conclusion}
We have  given a  simple sufficient condition to guarantee the secrecy for cryptographic
protocols which use pairing and symmetric and/or  asymmetric encryption. Secrecy is
ensured for an unbounded number of agents, nonces, sessions, without assuming any typing
of terms. Moreover, for a large class of protocols we provide an algorithm which
transforms a protocol into a  secure one w.r.t{.} secrecy and  preserves the "intended
goal" of the
original protocol. To our knowledge it is the first result of this type.\\
We have limited our work to protocols of depth at most two, which is reasonable from a
practical point of view. It seems that we could get rid of this restriction  easily, but
the proof would be more technical. A drawback of our sufficient condition is that the
systematic signature of messages with the private key of the sender increases the size of
the message. It would be better to replace $<\CS,\{\CS,m\}_{K_{priv}(A)}>$ with
$<\CS,m,\{H(\CS,m)\}_{K_{priv}(A)}>$ where $H$ is a hash function. We propose to extend
our study with more primitives, in particular with hash functions.
\bibliographystyle{alpha}
\bibliography{secrecy}

\end{document}